# Container Density Improvements with Dynamic Memory Extension using NAND Flash

Jan S. Rellermeyer
Delft University of Technology
j.s.rellermeyer@tudelft.nl

Maher Amer
Diablo Technologies
amer.maher@gmail.com

Richard Smutzer
Diablo Technologies
rasmutzer@gmail.com

Karthick Rajamani
IBM Research
karthick@us.ibm.com

## Abstract

While containers efficiently implement the idea of operating-system-level application virtualization, they are often insufficient to increase the server utilization to a desirable level. The reason is that in practice many containerized applications experience a limited amount of load while there are few containers with a high load. In such a scenario, the virtual memory management system can become the limiting factor to container density even though the working set of active containers would fit into main memory. In this paper, we describe and evaluate a system for transparently moving memory pages in and out of DRAM and to a NAND Flash medium which is attached through the memory bus. This technique, called Diablo Memory Expansion (DMX), operates on a prediction model and is able to relieve the pressure on the memory system. We present a benchmark for container density and show that even under an overall constant workload, adding additional containers adversely affects performance-critical applications in Docker. When using the DMX technology of the Memory1 system, however, the performance of the critical workload remains stable.

## 1 INTRODUCTION

Containerization is an ongoing and important trend within a broader effort to further improve server utilization. Instead of launching a dedicated virtual machine for every service and therefore having a separate OS kernel with all its overhead, multiple services can co-exist on the same operating system while still providing the level of isolation and resource management that is required in multi-tenant environments.

Systems like Docker [15] embraced container technology and built an ecosystem and tooling around it for easy, portable deployment of applications. This has led to many developers packaging their creations in containers. The widespread and growing popularity of containers today has resulted in most public clouds offering IaaS/PaaS to providing containers as a service in addition to virtual machines. Emerging application architectures using microservices [20] and cloud functions (functions-as-a-service or serverless computing [3]) also rely heavily on containers. While containers enable more efficient usage of infrastructure (relative to virtual machines) the sudden proliferation of them has created an unforeseen problem.

In IBM's cloud operation, we see many cases where a majority of the containers on a server are mostly inactive for an extended period of time while few containers show high activity. In order to keep the server utilization at a desired level, operators would like to go beyond machine partitioning and increase the density of container deployments by carefully over-committing the resources. However, the balancing act lies in increasing the density of containers by padding available capacity with lowly utilized container instances while not compromising the performance of critical containers.

Most of the experimental analysis of container systems has focused around the raw performance compared to either bare metal or virtual machine deployments [7], [19], [22]. Research on performance isolation in container systems has primarily considered storage [25], compute [26], and networking [27] while little attention has been paid to memory effects. In this paper, we present a benchmark for container density (§2) based on the idea of measuring the performance of a critical workload while adding an increasing number of moderately *noisy neighbors*. Our results for Docker (§3) show that even when the level of noise through mostly inactive tenant containers stays constant, the bare presence of more container instances can significantly impact the performance of a critical workload, primarily due to memory pressure. While the adverse effect on the performance due to an overall increase of page faults in presence of a higher memory access frequency is well studied and understood (e.g., [6], [11]), our setup performs the same number of memory accesses, just from a larger selection of processes. Under ideal conditions, the VMM page replacement system should be able to handle this gracefully since the number of hot pages remains essentially constant while only the number of cold pages increases. However, our benchmark shows a significant impact on the tail latency of the critical workload that runs concurrently with the mostly idle containers. Finally, we explore and evaluate a memory extension technique called *DMX* (Diablo Memory Expansion) in *Memory1* [14] by Diablo Technologies (in §4) that autonomously moves pages in and out of memory and to a DIMM-sized NAND Flash medium that is attached through the memory bus. By doing so, Memory1 can virtually extend the usable amount of memory at a much lower cost compared to the corresponding actual amount of DRAM. Our evaluation (§5) shows that DMX is able to increase the density of containers while leaving the critical workload unaffected.

## 2 CONTAINER DENSITY BENCHMARK

In order to assess the ability of the system to sustain a higher density of containers on the same machine we created a setup with a single performance-critical application that serves as a benchmark and a variable number of non-critical containers that perform periodic (but not benchmarked) activity. The intention of the setup is to measure the interference or non-interference of the non-critical with the critical workload.

The performance-critical workload is a deployment of the AcmeAir benchmark [1] which was developed by IBM Research. The benchmark simulates an online flight booking portal and determines the total number of transactions as well as min/max/avg latency of requests. In our setup we used the implementation based on node.js [23] as an application server running in one Docker container and MongoDB [18] as a data store running in a separate container. In addition, a second node.js instance handles the authentication of users so that the benchmark uses a total of three containers on the server side. Load to the system is generated by an Apache JMeter [10]-based multi-threaded client.

As *noise* workloads we use an Apache httpd webserver [9] (version 2.2) in a container which serves a series of large static image files. In order to stress the memory system, we explicitly cache the image files in the web servers using the mod_mem_cache module. We generate 500 images of 20 MiB size, 100 images of 80 MiB size, and 80 images of 200MiB. As a result, the memory footprint of a noise container is in the order of 34 GiB. We then program a client to randomly connect to one of the webserver instances and request an image file which it retrieves from memory. This can cause either 20, 80, or 200 MiB to be retrieved from RAM and potentially evicts pages from the working sets of other containers, including the AcmeAir instance. After a successful request the client sleeps for 150 ms and then randomly retrieves another image from a container instance.

The pressure on the system manifests itself primarily through the addition of more processes and therefore virtual address spaces. However, what makes containers particular is that they typically fulfill a single function (microservice architecture) and therefore all the processes within a container follow the same activation pattern and periods of inactivity. In this regard, the httpd webserver setup can be considered a typical example that is inactive for most of the time and, when activated through the arrival of a request, leads to a burst of memory accesses.

We conducted the experiments on an Inspur NF5180M4 2-socket server equipped with Intel Xeon E5-2660 v3 CPUs and 256 GiB of DRAM. We ran a Linux 3.10 kernel and Docker version 17.03.1-ce for which we disabled OOM kill in order to allow for over-committing the physical memory instead of having docker kill containers randomly to free up memory. An entire disk drive was dedicated to the swap partition.

## 3 EVALUATION OF DOCKER CONTAINER DENSITY

We repeatedly measured the throughput and latency of an AcmeAir setup while increasing the number of *noisy neighbors* in the form of httpd container instances with every iteration. In this experiment, we started with just one noise instance and scaled up to 49 instances, with a step size of two due to the length of the experiment. Under ideal conditions, we would expect the noise instances to have little to no impact on the AcmeAir performance since we keep the number of web requests and thereby the overall amount of memory traffic constant.

Figure 1a shows the transactions per second (TPS) that the booking system can process as a function of the httpd noise instances running on the same machine. The throughput remains relatively stable but then degrades quickly and noticeably at around 19 instances and the machine starts to swap. After 33 instances, the system even began to thrash and needed to be rebooted. It is important to note that the total noise workload does not effectively increase during the experiment since we do not change the number of clients, just the number of httpd servers serving the same amount of requests. In addition, we run the experiment in a contiguous fashion, meaning that after every iteration we keep the existing containers and add two new noise instances, which makes addition independent of the number of existing containers. What we observe is therefore the memory system not being agile enough to keep the hot pages in memory and paging cold pages out to the swap partition when containers are activated by receiving a request. We furthermore observed that the degradation is not primarily a function of the speed of the swap device since we saw comparable behavior on different types of media.

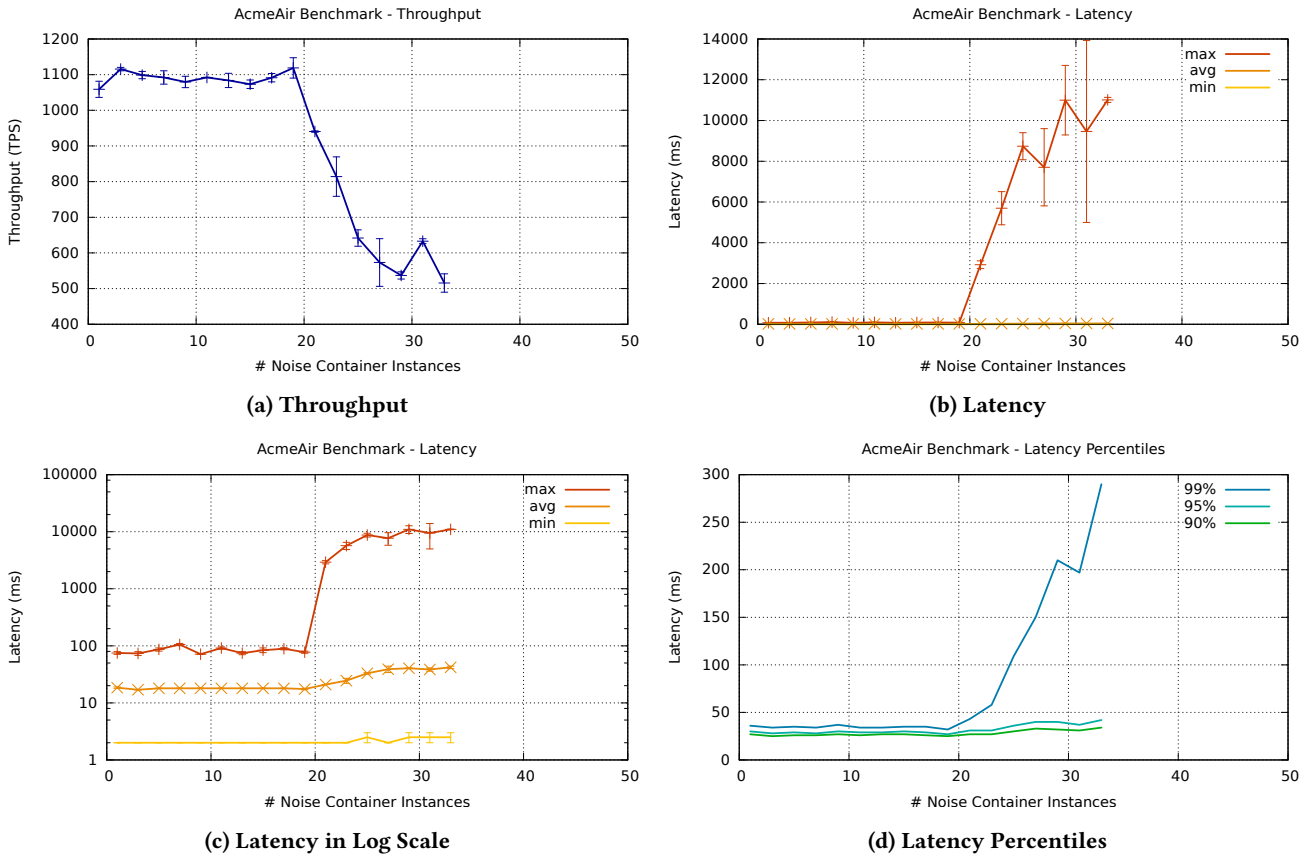

Figure 1: Throughput and Latency of the AcmeAir Workload Instance

A look at the latency (Figure 1b) suggests that the throughput is suffering because of a spike in the tail latency [5]. While the minimum is hardly affected, the maximum latency for requests to get processed increases by more than three orders of magnitude. The log-scale representation (Figure 1c) shows that the average latency increases by a factor of two but not super-linear as the max. An in-depth analysis of the latency percentiles in Figure 1d confirms this hypothesis and shows that even the 95th percentile is hardly affected and the spike in latency is only visible in the 99th percentile.

In summary, the experimental results against a conventional server system appear sound as a baseline for container density and show that an increasing number of noisy neighbor can negatively affect a workload even when the total amount of load remains constant, which matches our experience from production systems. The performance degradation is significant (around 50% after hitting the saturation point of the VMM) and the effect manifests itself over-proportionally in the tail latency.

## 4 DYNAMIC MEMORY EXTENSION

In traditional forms of memory management, both the mapping and the migration of pages between the different tiers of memory and storage is under the control of the operating system. With the mechanisms, operating systems furthermore implement policies and heuristics to decide on an optimal placement of pages. As a result, the presence of multiple memory tiers is fully transparent to applications and jointly create the illusion of an (almost) unbound virtual memory address space.

Memory1, in contrast, is a server memory extension product developed by Diablo Technologies that marries a fast storage device with an external memory manager in software that utilizes a prediction model to manage page migration and placement more efficiently. In contrast to, e.g., the Linux Swap system which relies fully on a page fault handler and reacts the memory pressure, DMX is continually running and monitoring memory traffic from the time an application makes memory allocation request until the process terminates. Throughout that time, DMX monitors memory traffic (without being in the memory path) and learns memory pattern behavior even without page faults.

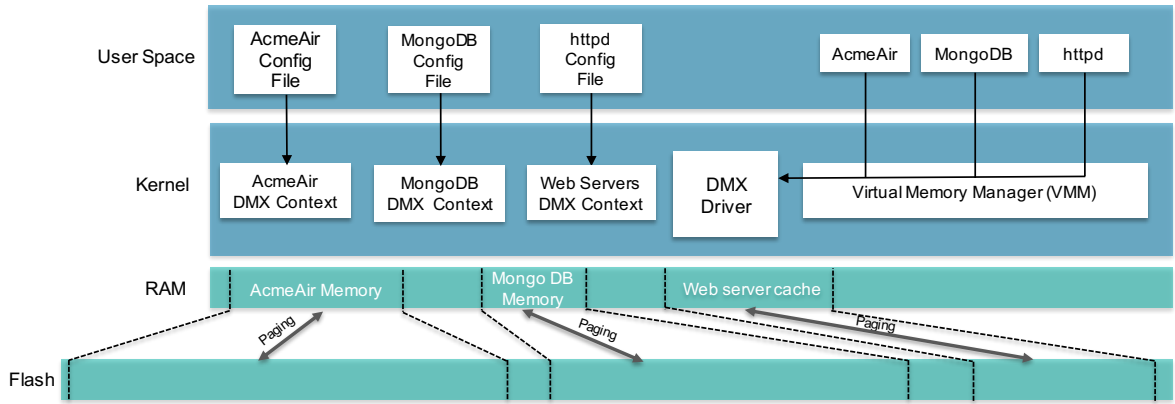

Figure 2: Structure of the Memory1 DMX System

Memory1 consists of a software component called DMX SW and a hardware component called DMX module which is a NAND-flash module that plugs into the DDR-4 memory channel and provides high bandwidth, low latency access to flash devices mounted on the module. To be able to use DMX modules, server BIOS/UEFI modifications are required. Beyond that, however, Memory1 is transparent to the operating system and semi-transparent to applications in that it requires the user to configure which processes should receive memory extension.

DMX SW is a loadable kernel module driver that connects onto the OS Virtual Memory Manager (VMM) and can be considered an extension of the operating system's VMM. DMX SW supports major Linux distributions (RedHat, Ubuntu, etc) and does not require any changes to the OS or applications. The user can select which application(s) are to be managed by DMX via a configuration file. By means of a pre-load library in the user space, the DMX kernel driver intercepts and services all memory requests generated by the selected application (malloc, page fault, etc). DMX creates a *Memory Context* for each selected application running on the server and carves out a *dynamic* portion of the server's physical memory (i.e. DRAM) to be used as front-end cache for that application (Figure 2). It is dynamic because the size of that cache will dynamically vary depending on various factors, the primary being the degree of memory activity of the specific application. Since DMX is servicing all memory requests of the selected application, it is aware of its memory access activities as well as its access patterns. The memory access patterns are used to feed the tiering and prediction algorithms in the application *Memory Context* and the memory activities will be used to decide on the cache size allocated for this application. A dormant application or one that is only doing compute/processing on small region of its memory will consume a small amount of cache. If the total needed cache sizes of all running applications exceeds the total amount of physical memory available in the system, the tiering and predictions algorithms will make sure the *appropriate* data remains in physical memory.

Although DMX SW was originally developed to only work with DMX modules, its current version is capable of utilizing industry-standard NVMe-PCI SSDs as the flash module to be used as a paging device, and hence eliminate the required BIOS/UEFI changes or limitations in server hardware. Memory1 was designed to be hardware agnostic and accordingly, DMX SW is split into two distinct parts: Data Management and Media Management. The Data Management element is where all tiering and prediction algorithms are implemented. It also contains the interface that attaches DMX SW onto the operating system stack. Data Management interfaces with Media Management by generating block requests (pages to be evicted/written and pages to be prefetched/read). The Media Management element is responsible for translating these block requests into standard IO requests and has built in intelligence to maximize the performance of the paging media (traffic shaping). Having Memory1 run with alternative flash media (NVMe SSDs vs. DMX modules) only requires changes to the Media Management resulting in large degree of hardware compatibility and scalability.

Traditional virtual memory management is primarily reactive. The operating system reacts to page faults and then loads pages back into main memory, applying a heuristic replacement policy in order to reduce the likelihood of future page faults. Linux, e.g., uses a simple LRU queue to determine data hotness and hence make decisions on what stays in memory and what should not. Memory1, in contrast, uses a prediction model to proactively migrate pages in and out of main memory and therefore eliminates page faults that would trigger the operating system's VMM. It is based on a complex multi-queue algorithm for data tiering and relies on machine learning techniques to creates statistical model of all high frequency pages to enable accurate prediction of which data to prefetch from Flash. By doing that, paging device latency is amortized across multiple page faults and hence reducing its negative impact. Furthermore, Linux Swap works on the whole data as aggregate entity. As a result, it does not distinguish which data belongs to which process or container. That renders it inefficient when containers go from active to inactive. DMX on the other hand manages each container data independently and should therefore be able to efficiently detect when container go inactive and back to active and react accordingly by prefetching the associated data. In the following experiment, we explore to which extent DMX is able to increase the density of containers in practice on a single machine by reducing the overhead of the Linux VMM.

## 5 CONTAINER DENSITY WITH DYNAMIC MEMORY EXTENSION

We ran the same experiment as in §3 but this time with the DMX software enabled. Every container was given a configuration file so that DMX could manage it. On the hardware side, 12 of the memory slots were populated with 128 GiB Memory1 NAND Flash DIMMs for a total of 2 TiB of extended memory. The firmware version was still under active development at the time of the experiment but the changes necessary to support a Docker container environment were released to the general public with version 1.3 of the DMX SW.

Since the total amount of memory used at any time does not change during the previous experiment it is clear that the system is not overloaded in the strict sense but instead what we observe is the inability of the virtual memory system to keep up with the growing number of tenant containers (i.e., processes). This should give Memory1 an opportunity to relieve the memory system by moving pages out of DRAM (and thereby out of the realm of the operating system's virtual memory manager) and into NAND Flash.

In order to validate if dynamic memory extension is able to increase the density of containers on servers we repeated the experiment, this time with Memory 1 DMX enabled. In our configuration, the amount of Flash is eight times the size of the DRAM. Since the technology is based on predictions, a moderate amount of overhead can be expected since our requests are randomized and some pages need to be fetched from the slower Flash medium within the critical path of the AcmeAir workload.

The throughput of AcmeAir, shown in Figure 3a, remained almost constant over the entire duration of the run. In absolute terms, it is about 10% lower than with no memory extension but the system can now sustain the 49 noise instances without a noticeable performance degradation. The latency graph (Figure 3b) shows that there is once again a slight overhead that the system introduces when compared to the case without memory extension but the latency remains largely unaffected regardless of the number of noise instances running. A look at the log-scale latency (Figure 4a) and latency percentiles (Figure 4b) reveal that the slightly higher average latency is mostly due to an increased degree of jitter. The maximum latency is higher and more volatile due to the impact of mis-predictions of the DMX technology and consequently page loads from the Flash device. The 90th percentile and 95th percentile are close to identical to the non-DMX case for a low enough amount of noise but the 99th percentile is slightly higher.

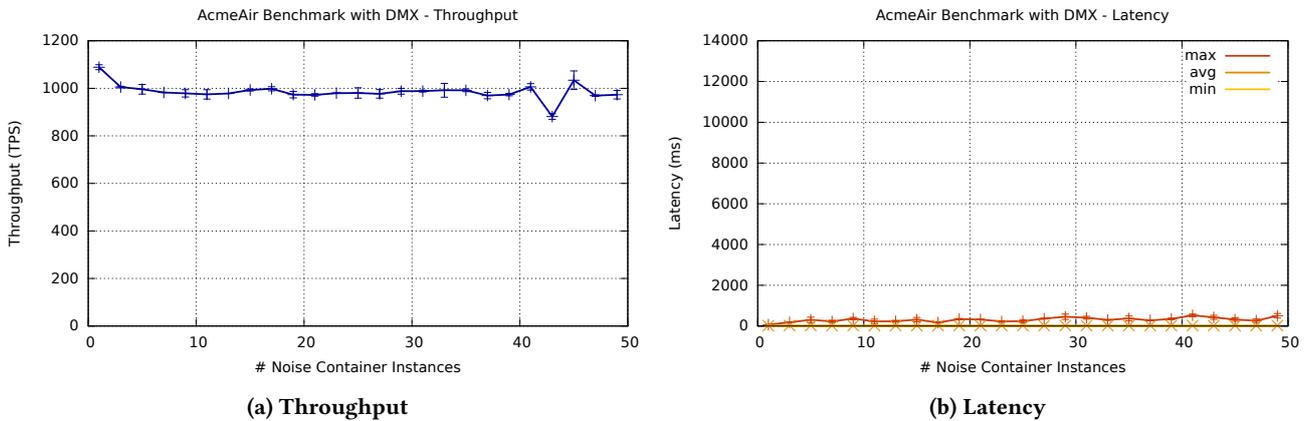

(a) Throughput    (b) Latency

Figure 3: Throughput and Latency of the AcmeAir Workload Instance with DMX

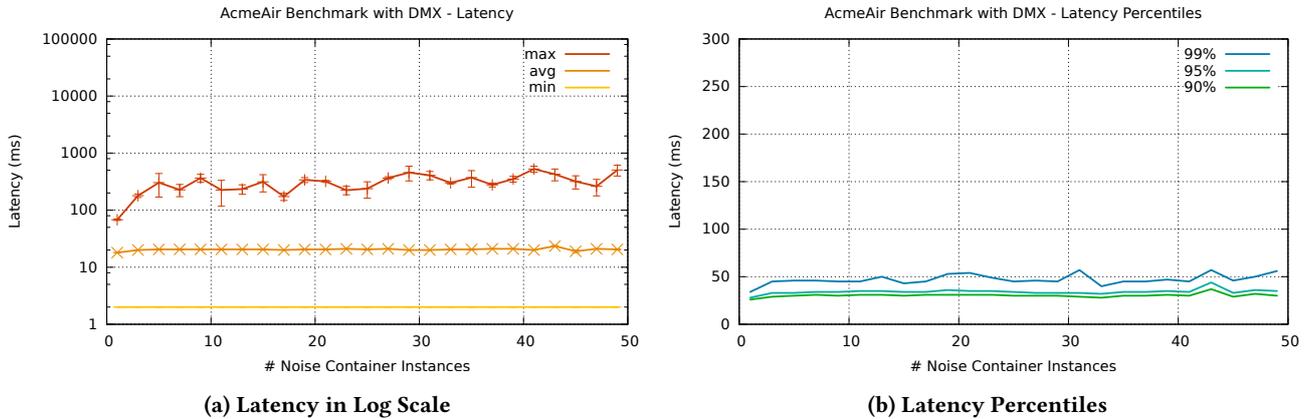

(a) Latency in Log Scale

(b) Latency Percentiles

Figure 4: Latency of the AcmeAir Workload Instance with DMX

## 6 RELATED WORK

We are not the first to explore Flash/SSDs as a way to extend main memory. Early work primarily revolved around optimizing the existing Linux kernel swap mechanism for traditional Flash storage. FASS [12] aimed at moving the flash translation layer into the kernel swap mechanism for better efficiency and durability. LOBI [13] optimized the way how pages are stored on the medium by using a log-structured swap-out and a block-aligned swap-in. FlashVM [21] presented more fundamental changes to the Linux virtual memory system to better support Flash for paging. This includes prefetching of pages on the read path, throttling and zero-page sharing on the write path, and garbage collection.

Mogul et al. proposed to combine Flash and DRAM into a single package and elaborated on the required operating system support for such forms of hybrid memory [17]. Their main conclusion is to use the DRAM portion for buffering. Memory1 achieves a comparable effect through their prediction model and by using DRAM as a front-end cache for Flash.

SSDAlloc [2] is a hybrid DRAM and flash memory manager which allows applications to extend their usable amount of memory in a semi-transparent way. It replaces *malloc* and utilizes the Flash medium as a log-structured object store. In comparison, Memory1 is fully transparent and operates on the granularity of pages

Intel has recently announced a byte-addressable storage-class memory solution [8] with XPoint [4]. While it shares the motivation with Memory1 to offer a middle ground between DRAM and storage in terms of price and latency, the technology is not yet available in the DIMM form factor. However, when available, it could be used with the DMX software for dynamic memory extension without requiring changes to DMX.

Vilayannur et al. [24] explored the challenges of traditional page replacement algorithms for scientific applications that tend to access memory in a cyclic manner. Their conclusion was that as soon as the working set became slightly larger than the available main memory, most algorithms tend to evict exactly the pages that are likely to be accessed next. In response, they proposed the use of predictive page replacement techniques. Memory1 uses a predictive algorithm and combines it with a fast and high-bandwidth secondary storage device. While our experimental setup for the noise containers is not cyclic in nature but performs random access of larger blocks of memory and sequential access within the blocks, it unveiled similar problems with the standard Linux VMM as the authors had observed.

Mesnier et al. proposed Differentiated Storage Systems [16] to address the challenge of the single policy approach of most operating system when it comes to providing storage to applications. Their work exposes an API through which the application can provide a classification of storage blocks which is then mapped to a custom policy. One of the intended applications is caching of blocks in the buffer cache. DMX shares the motivation for differentiating services per application but does so for memory as opposed to storage. Furthermore, it provides differentiation in a manner that is transparent to both the application and the operating system, without the need for API changes while the implementation equivalent to Differentiated Storage Systems would require the application to provide classification with every memory allocation request.

## 7 CONCLUSIONS

We have presented a benchmark for container systems that is able to show the correlation between container density of co-located low priority containers and tail latency of a performance-critical container workload. In our setup, we could observe a 2x reduction of throughput and increase of average latency while the tail latency increases by several orders of magnitude. Memory1's dynamic memory extension technique, however, has proven to be capable of increasing the density of containers and mitigating the problem of paging in dense container deployments. With DMX enabled, the same system can support more than twice as many co-located low priority containers without significant performance degradation by predicting the pages that are likely accessed when a container becomes active. Through our experiments, we have shown that dynamic memory extension with Flash DIMMs is a viable option to increase the density in container deployments at a lower price point than increasing the amount of main memory by adding more DRAM.

## 8 AVAILABILITY OF THE SETUP

The scripts to build and run the experiment are available on GitHub (https://github.com/rellermeyer/container_scale) and are provided under the Apache 2 license.